\documentclass[12pt,a4paper]{article}
\usepackage[british]{babel}

\usepackage[a4paper,top=2cm,bottom=3cm,left=2.5cm,right=2.5cm,marginparwidth=1.75cm]{geometry}

\pdfoutput=1
\usepackage{amssymb,amsmath,bm}
\usepackage{graphicx}
\usepackage{subcaption}
\graphicspath{ {Images} }
\usepackage[final]{hyperref}

\usepackage{float}
\usepackage[usenames,dvipsnames]{color}
\usepackage[normalem]{ulem} 
\renewcommand\sout{\bgroup \color[rgb]{0.55,0.00,0.99} \ULdepth=-.5ex \ULset}

\begin{document}


\begin{center}
{\Large Constraining the proton transverse partonic distribution through coherent diffractive $J/\psi$ production at HERA within a static Gaussian hot spot model} \\[5ex]

{\large Muhammad Raihannafi Fadhel} \\
{\small
Department of Physics, Universitas Gadjah Mada, BLS 21, Yogyakarta, Indonesia \\
muhammad.raihannafi.fadhel@mail.ugm.ac.id} \\ [1.2em]

{\large Chalis Setyadi} \\
{\small
Department of Physics, Universitas Gadjah Mada, BLS 21, Yogyakarta, Indonesia \\
chalis@mail.ugm.ac.id
} \\[1.2em]

\end{center}

\begin{abstract}
We investigate the transverse parton distribution of the proton through the $t$-dependence of the coherent $J/\psi$ differential cross section extracted from HERA measurements in the small-$x$ regime. Employing a simple static Gaussian hot spot model inspired by the large-$x$ three–valence-quark picture of the proton, we introduce three geometric degrees of freedom of each hot spot: (1) the Gaussian width of each hot spot, (2) the spatial distance of the valence quarks from the proton center, and (3) their azimuthal orientation. In the simplest implementation, the valence quarks are assumed to form a symmetric triangular configuration. We find that variations in these geometric structures significantly influence the $t$-slope of the coherent cross section. In particular, the rotational degree of freedom strongly affects the cross section at $t>0.5\,\mathrm{GeV}^2$, while the small-$t$ region remains largely insensitive to variations in this parameter.
\end{abstract}


\section{Introduction} \label{Intro}

Understanding the internal structure of the proton remains a central question in high-energy QCD. Deep inelastic scattering (DIS) experiments, particularly those conducted by the H1 and ZEUS collaborations at HERA \cite{H1:2005dtp,H1:2003ksk,H1:2013okq,ZEUS:1999ptu,ZEUS:2004yeh} have provided an extensive set of lepton–proton collision data. These measurement data have been analyzed using a variety of phenomenological approaches, see, for instance \cite{Golec-Biernat:1998zce,Kowalski:2003hm,Marquet:2007qa, Garcia-Montero:2025ekv, Hentschinski:2020yfm, Flett:2025chf, Kou:2025gwm, Goloskokov:2024egn,Penttala:2024hvp}. 
Insights from collinear parton distribution functions (PDFs) \cite{Dulat:2015mca,NNPDF:2017mvq} indicate that, while at low energies the proton is primarily composed of three valence quarks bound by gluons, it exhibits a far more complex and fluctuating substructure at high energies, consistent with expectations from the Color Glass Condensate (CGC) framework\cite{Gelis:2010nm,Iancu:2000hn,Weigert:2005us,Venugopalan:2005mg}.

In order to describe high-energy collisions, the dipole picture ~\cite{Mueller:1999yb} is a well-known approach. Over the past two decades, this framework has been employed in many models and has been able to describe a wide range of HERA data. One of its most appealing features is the transparent physical interpretation it offers at small Bjorken-$x$. In this picture, a virtual photon emitted by a lepton fluctuates into a dipole (a quark–antiquark pair) which subsequently interacts with the target proton. Even in its simplest implementations, the dipole formalism successfully describes diffractive virtual photon–proton cross sections, and its extensions are widely used to parametrize HERA data. At sufficiently small $x$, parton densities inside the proton become so large that non-linear QCD dynamics emerge. In this saturation regime, collinear factorization and linear evolution are no longer valid, and the proton is better described as a state of strongly interacting, high-occupancy gluonic matter known as the CGC. The dipole model was originally developed precisely to capture these saturation phenomena.

Phenomenologically, the essential challenge is to connect experimental observables to dipole–target interactions.
In modeling the spatial structure of the proton, several phenomenological models have been proposed that incorporate the impact-parameter dependence of the partonic distribution \cite{Kowalski:2003hm,Kowalski:2006hc,Rybczynski:2013mla,Mantysaari:2016jaz}. Among these, the hot spot model has emerged as a physically motivated description. 
It assumes that large-$x$ valence quarks act as centers from which small-$x$ gluon fields radiate, giving rise to localized “hot spots” in the transverse plane.
This model naturally incorporates event-by-event fluctuations, which are essential for describing coherent and incoherent processes and align well with phenomenological evidence from HERA and recent theoretical developments~\cite{Mantysaari:2016jaz}.

In this work, our goal is to constrain various parameters of a simplified version of the hot spot model—specifically, the static hot spot model in which fluctuation effects are ignored. We study how parameters related to the hot spot configuration can be probed through measurements of the exclusive $J/\psi$ production cross section in HERA data, particularly when the positions of the hot spots are varied in both rotational and spatial directions. Additionally, we examine the effect of the hot spot size on the cross section. For this purpose, we use the simplified form of the hot spot model proposed by \cite{Mantysaari:2016jaz}. The structure of this paper is as follows. We begin in Section \ref{theory} with a short review of the dipole picture for exclusive vector meson production, along with the photon and vector meson wave function overlap needed for the scattering amplitude. Section \ref{hotspotmodel} introduces the hot spot model along with the free parameters. Finally, in Section \ref{results}, we discuss the comparison of our model with HERA data on exclusive $J/\psi$ production cross sections and present our conclusions in Section \ref{conclusions}.

\section{Exclusive Coherent $J/\psi$ Production in the Dipole Picture} \label{theory}

The dipole picture is often employed to investigate high energy lepton–proton collisions \cite{Mueller:1994jq,Chen:1995pa}. In this picture, exclusive coherent vector-meson production in $ep$ collisions can be viewed as follows: the electron emits a virtual photons which later fluctuates into color dipoles in the form of quark and anti-quark pairs, and interact with a single parton inside the proton. Subsequently, the proton and the dipole are scattered, with the proton remains intact \cite{Golec-Biernat:1998zce}. Meanwhile, the dipole forms a meson as the final state. 

The differential cross section for vector meson production in photon–proton scattering can be written in the form of \cite{Kowalski:2006hc} 
\begin{equation}
    \frac{d\sigma^{\gamma^{(*)}p\rightarrow Vp}_{T,L}}{dt} = \frac{1}{16\pi} \left|{\mathcal{A}^{\gamma^{(*)}p\rightarrow Vp}_{T,L}}\right|^2 \label{dsigmadete}
\end{equation}
where $\mathcal{A}^{\gamma^{(*)}p\rightarrow Vp}_{T,L}$ denotes the scattering amplitude with ($T$) and ($L$) represents the transverse  and longitudinal polarization states, respectively. Explicitly, the scattering amplitude is
\begin{equation}
    \mathcal{A}^{\gamma^{(*)}p\rightarrow Vp}_{T,L} = 2i \int d^2\bm{r}_{\perp} \int_0^1 \frac{dz}{4\pi} (\Psi_V^* \Psi)_{T,L} \int d^2\bm{b}_{\perp} e^{-i\left(  \bm{b}_{\perp} - \left(\frac{1}{2}-z\right)\bm{r}_{\perp}\right)\cdot\Delta_{\perp}} \hat{\sigma}(x,\bm{r}_{\perp},\bm{b}_{\perp}).\label{elasticscatteringamplitude}
\end{equation}
Here, $\bm{b}_{\perp}$ denotes an impact parameter of the dipole with respect to the target (proton) center, $\bm{r}_{\perp}$ represent quark and anti-quark transverse separation distance, $\Delta_{\perp}=\sqrt{-t}$ is the transverse momentum transfer of the proton, $Q^2$ is the photon virtuality, and the longitudinal momentum fraction carried by quark is denoted by $z$ (an anti-quark carried a fractional momentum with the magnitude of $1-z$, respectively). Furthermore, $(\Psi_V^* \Psi)_{T,L}$ denotes the wave functions overlap between virtual photon and the vector meson, and $\hat{\sigma}(x,\bm{r}_{\perp},\bm{b}_{\perp})$ is a dipole scattering amplitude. 
The photon wave function can be calculated perturbatively, see for example \cite{Lepage:1980fj,Dosch:1996ss}. However, the vector meson wave function is more complicated, hence several phenomenological models have been proposed. In this study we choose the most commonly used wave function, the Boosted Gaussian (BG) \cite{Kowalski:2006hc}. The overlap of the photon and meson wave functions for transverse polarization can be written as \cite{Forshaw:2003ki}:
\begin{align}
    (\Psi^*_V\Psi)_T&=\hat{e}_fe\frac{N_c}{\pi z(1-z)}\left[m_f^2K_0(\epsilon \bm{r}_{\perp})-(z^2+(1-z)^2)\epsilon K_1(\epsilon r)\partial_{\bm{r}_{\perp}}\right]\phi_T(\bm{r}_{\perp},z)\nonumber \\
    (\Psi^*_V\Psi)_L&=\hat{e}_fe\frac{N_c}{\pi}2Qz(1-z)K_0(\epsilon \bm{r}_{\perp})\left[M_V+\frac{m_f^2-\nabla^2_{\bm{r}_{\perp}}}{M_Vz(1-z)}\right]\phi_L(\bm{r}_{\perp},z), \label{PSIVGTL}
\end{align}
where $\phi_L(\bm{r}_{\perp},z)$ are the scalar part of the vector meson wavefunction. Here $e=\sqrt{4\pi \alpha_{em}}$ denotes the electric charge, $\epsilon^2=z(1-z)Q^2+m_f^2$, and $N_c = 3$ represent the number of color charge on the standard model. The flavor dependence $f$ appears through the value of the quark charge $e_f$ and the quark mass $m_f$. The scalar part for the BG model is given by \cite{Nemchik:1994fp,Nemchik:1996cw,Forshaw:2003ki,Kowalski:2006hc}:
\begin{equation}
    \phi_{T,L}(\bm{r}_{\perp},z) = \mathcal{N}_{T,L}z(1-z)\exp\left(-\frac{2z(1-z)\bm{r}_{\perp}^2}{\mathcal{R}^2}-\frac{m_f^2\mathcal{R}^2}{8z(1-z)}+\frac{m_f^2\mathcal{R}^2}{2}\right). \label{BGModel}
\end{equation}
The free parameters $\mathcal{N}_{T,L}$ and $\mathcal{R}$ are determined based on the wave function normalization and the leptonic decay width of the meson. For $J/\psi$, we use: $\mathcal{N}_T = 0.578$, $\mathcal{N}_L = 0.575$, $\mathcal{R}^2 = 2.3 \; \text{GeV}^{-2}$, $M_V = 3.097$ GeV, and $m_f=1.4$ GeV \cite{Kowalski:2006hc}. Other numerical values of these parameters can also be found in \cite{Mantysaari:2018nng}.

For the dipole scattering amplitude, we used a model developed based on the McLerran–Venugopalan (MV) model \cite{McLerran:1993ka,McLerran:1993ni,McLerran:1994vd}, which was later modified to incorporate the $x$ dependence, which allows us to include saturation effects at small $x$, for example \cite{Gelis:2010nm,Iancu:2000hn,Mueller:1999yb,Ferreiro:2001qy,Iancu:2003ge}. The longitudinal momentum fraction of the parton $x$ is defined as $x = (M_V^2 + Q^2 )/(W^2+Q^2)$, with $W$ is the photon–proton center of mass energy \cite{Martin:1997wy,Martin:1999wb,Watt:2007nr}. The energy evolution of the saturation scale is incorporated through the $x$-dependent Golec-Biernat–W\"{u}sthoff (GBW) parameterization \cite{Golec-Biernat:1998zce,Golec-Biernat:1999qor}, which has been shown to accurately describe DIS measurements at HERA in the small-$x$ limit \cite{Stasto:2000er}. The dipole scattering amplitude is therefore written as \cite{McLerran:1993ka,McLerran:1993ni,McLerran:1994vd}:
\begin{equation}
    \hat{\sigma}(x,\bm{r}_{\perp},\bm{b}_{\perp})=1-\exp\left(-\frac{1}{4}\bm{r}_{\perp}^2Q_s^2(x,\bm{r}_\perp,\bm{b}_\perp)\ln{\left[\frac{1}{\bm{r}_{\perp}^2\Lambda_{\text{QCD}}^2}+e\right]}\right),
\end{equation}
while the saturation scale is parameterized by \cite{Boer:2023mip}
\begin{equation}
    Q_s^2(x,\bm{r}_\perp,\bm{b}_\perp) = \bar{\chi}\left(\frac{x_0}{x}\right)^{\lambda}\alpha_s(\bm{r}_\perp)T_H(\bm{b}_\perp), \label{saturationscale}
\end{equation}
where $x_0=3\times10^{-4}$ and $\lambda=0.22$.
In our model, we use free parameter $\bar{\chi}$ to control the overall value of the differential cross section.
The proton profile $T_H(\bm{b_\perp})$ which describes the transverse distribution of color charges in the proton, is the only term where the impact parameter enters the dipole cross section. The impact parameter $\bm{b}_\perp$ is Fourier conjugate to the transverse momentum transfer $\Delta_\perp$, and therefore this profile dominates the slope of the differential cross section $d\sigma/dt$. Our study thus focuses on modeling this transverse color charge distribution.

The final ingredient in the cross section is the running strong coupling, taken in the form \cite{Mantysaari:2022sux}:
\begin{equation}
    \alpha_s(\bm{r}_\perp)=\frac{12\pi}{(33-2N_F)\ln\left[\left(\frac{\mu_0^2}{\Lambda_{\text{QCD}}^2}\right)^{\frac{1}{c}}+\left(\frac{4}{\bm{r}_\perp^2 \Lambda_{\text{QCD}}^2}\right)^{\frac{1}{c}}\right]^c},
\end{equation}
where the parameters are chosen to be $\mu_0=0.28$ GeV, $c=0.2$, and $\Lambda_{\text{QCD}}=0.040$ GeV as used in \cite{Mantysaari:2022sux}.
The numerical value of $\Lambda_{\text{QCD}}$ appears to be quite small, which is obtained from a fit of the model to data and should therefore be interpreted as an effective fit parameter rather than the fundamental QCD scale. With this choice, the model studied in \cite{Mantysaari:2022sux} provides a good description of the coherent $J/\psi$ differential cross section and yields a reasonable value of the strong coupling $\alpha_s=0.14$ at the typical scale $\bm{r}_\perp \sim1/M_{J/\psi}$.

\section{Static Gaussian Hot Spot Model} \label{hotspotmodel}

\begin{figure}[h]
\centering
\includegraphics[width=0.48\textwidth]{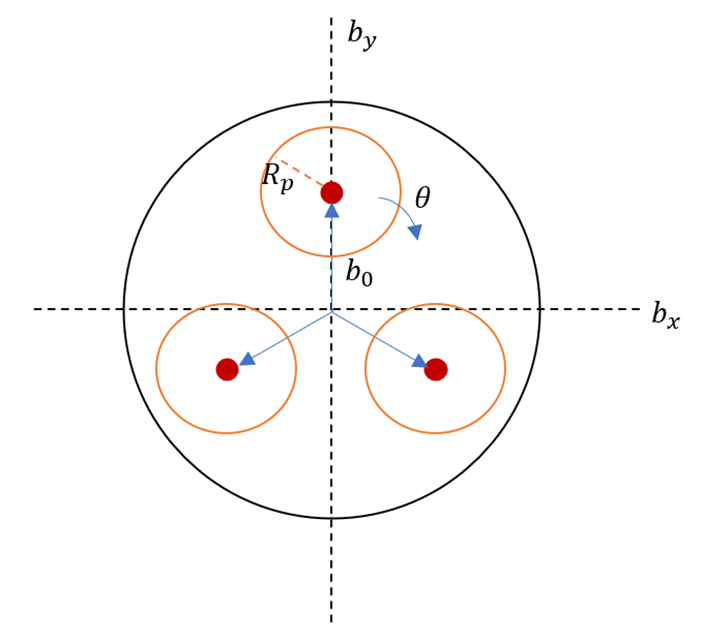}
\caption{The configuration of three valence quarks in the proton, forming an equilateral triangle.}
\label{ilustrasihotspot}
\end{figure}

In order to represent the partonic distribution of the proton, several analyses using hot spot models have been performed, see, for example\cite{Albacete:2016pmp,Albacete:2017ajt,Mantysaari:2016jaz,Mantysaari:2016ykx,Mantysaari:2017dwh,Cepila:2016uku,Cepila:2017nef,Mantysaari:2018zdd,Traini:2018hxd,Kumar:2021zbn,Kumar:2022aly,Demirci:2021kya,Demirci:2022wuy,Mantysaari:2022ffw}.
In general, a hot spot model is based on a large-$x$ proton picture, which consists of three valence quarks that can be thought of as sources of small-$x$ partons emitted around the constituent quarks \cite{Schlichting:2014ipa}. The color charge is then distributed around each valence quark. In our model, the color charge is distributed according to a Gaussian profile, with width parameter $R_p$ and can be explicitly expressed as:
\begin{equation}
    T_H(\bm{b_\perp})=\frac{1}{2\pi R_p^2N_q}\sum_{i=1}^{N_q=3}\exp\left(\frac{\bm{b_\perp}-\bm{b}_{i\perp}}{2R_p^2}\right).
\end{equation}
Furthermore, for simplicity, we assume that the valence quarks are located at the same radial distance $b_0$ from the center of the proton, forming an equilateral triangle as illustrated in Fig. \ref{ilustrasihotspot}.

In our static Gaussian hot-spot model, we vary three degrees of freedom: the Gaussian width parameter of each hot spot $R_p$, the radial distance of the constituent quarks from the proton center $b_0$, and the azimuthal orientation of the quark configuration $\theta$ which undergoes a rigid rotation about the $b_z$ axis in the clockwise direction. As our initial reference for the azimuthal angle, we use the configuration shown in Fig. \ref{ilustrasihotspot}, corresponding to $\theta=0$.
Since we consider a high energy process, we assume that Lorentz contraction renders the proton a two dimensional thin disk, and therefore only its transverse distribution is relevant.
Explicitly, the initial transverse position vectors of the valence quarks are given by
\begin{eqnarray}
\bm{b}_{1\perp}=\left(0,b_0\right), \quad, \bm{b}_{2\perp}=\left(\frac{\sqrt3 b_0}{2},-\frac{b_0}{2}\right), \quad \textrm{and}, \quad
\bm{b}_{3\perp}=\left(-\frac{\sqrt3 b_0}{2},-\frac{b_0}{2}\right).
\end{eqnarray}

\section{Numerical Results} \label{results}

As a standard practice, the cross section for the $\gamma^{(*)}p\rightarrow Vp$ process defined in Eq. \eqref{dsigmadete} is evaluated assuming that the scattering amplitude $\mathcal{A}^{\gamma^{(*)}p\rightarrow Vp}_{T,L}$ is purely imaginary. To account for the contribution from the real part of the amplitude, one typically applies the optical theorem, which leads to a correction factor $(1+\beta_{T,L}^2)$. This phenomenological correction can be obtained from the following relation \cite{Martin:1999wb,Shuvaev:1999ce}
\begin{align}
    \beta_{T,L}=\tan\left(\frac{\pi \lambda_{T,L}}{2}\right), \;\;\;\;\;\;\; \lambda_{T,L}=\frac{\partial \ln\left(\mathcal{A}^{\gamma^{(*)}p\rightarrow Vp}_{T,L}\right)}{\partial \ln \frac{1}{x}}.
\end{align}
Another correction to the cross section that must be taken into account is the skewedness effect. Since the dipole scattering amplitude usually assumes that the two exchanged gluons carry the same longitudinal momentum fraction of the proton, the skewedness correction accounts for the possible difference between these momentum fractions and is introduced through the factor defined as \cite{Shuvaev:1999ce}:
\begin{align}
    R_g=\frac{2^{2\lambda+3}}{\sqrt{\pi}}\frac{\Gamma(\lambda+\frac{5}{2})}{\Gamma(\lambda+4)}.
\end{align}
After incorporating the real-part and skewedness corrections, the differential cross section is given by
\begin{equation}
    \frac{d\sigma^{\gamma^{(*)}p\rightarrow Vp}_{T,L}}{dt} = \frac{1}{16\pi} \left|{\mathcal{A}^{\gamma^{(*)}p\rightarrow Vp}_{T,L}}\right|^2(1+\beta_{T,L}^2)R_g^2. \label{dsigmadetecorr}
\end{equation}
Several phenomenological studies \cite{Martin:1999wb,Toll:2012mb,Mantysaari:2016jaz} indicate that these corrections are typically of order $10\%-25\%$ for the real-part correction and $40\%-70\%$ for the skewedness effect. We will investigate the effect of each correction on the cross section in this work. In this study, we define the ratio of the differential cross sections with and without these corrections as
\begin{equation}
    R_{\text{corr}}=\frac{|\mathcal{A}_{T,L}|^2(1+\beta_{T,L}^2)R_g^2}{|\mathcal{A}_{T,L}|^2}. \label{rcorr}
\end{equation}

For the numerical calculation, we fix $R_p=0.165$ fm, following the value obtained from fits to coherent $J/\psi$ production at HERA in Refs. \cite{Mantysaari:2016jaz,Demirci:2022wuy}.
Using the hot spot proton profile described in Section \ref{hotspotmodel}, we impose the additional constraint $b_0>2R_p$ to minimize the overlap between the Gaussian color-charge distributions of each hot spot. Hence, the parameter $b_0$ directly controls the geometry of the proton. To keep the proton profile realistic and consistent with the phenomenological proton size, we therefore require a second constraint, $b_0+R_p\le0.6$ fm. Within these constraints we find a preferred value $b_0=0.43$ fm, which yields a reasonable slope for the $t$-dependent differential cross section compared with the HERA data over a wide range of photon virtualities $Q^2$ \cite{H1:2005dtp,ZEUS:2004yeh}, while being consistent with phenomenological expectations for the proton size. The remaining free parameters are obtained by rotating the positions of the valence quarks and varying $\bar{\chi}$. This procedure yields an overall good description for $\theta=0.12$ and $\bar{\chi}=28.8$.

\begin{figure}[h]
\centering

\begin{subfigure}{0.32\textwidth}
    \centering
    \includegraphics[width=\textwidth]{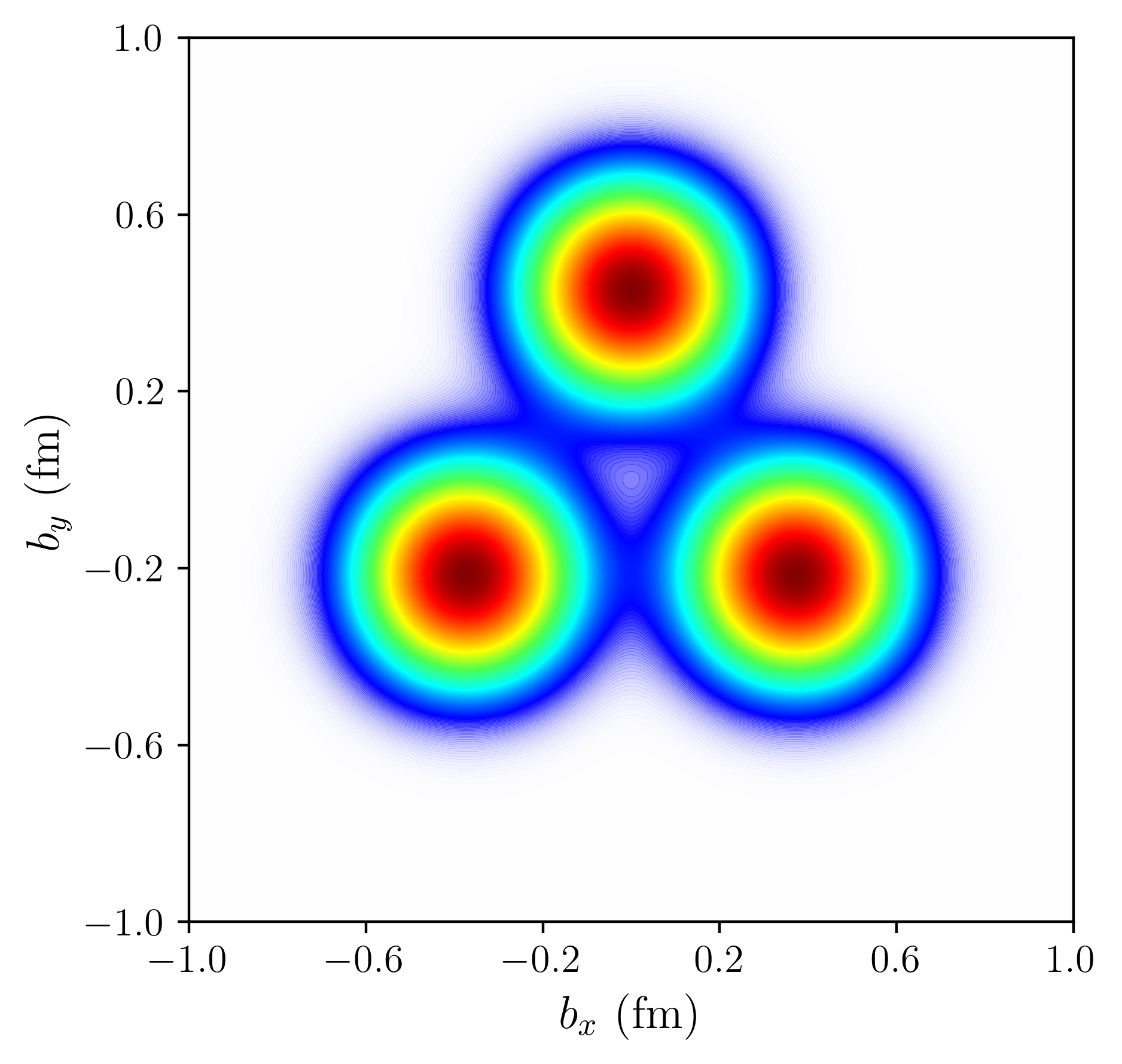}
    \caption{}
\end{subfigure}\hfill
\begin{subfigure}{0.32\textwidth}
    \centering
    \includegraphics[width=\textwidth]{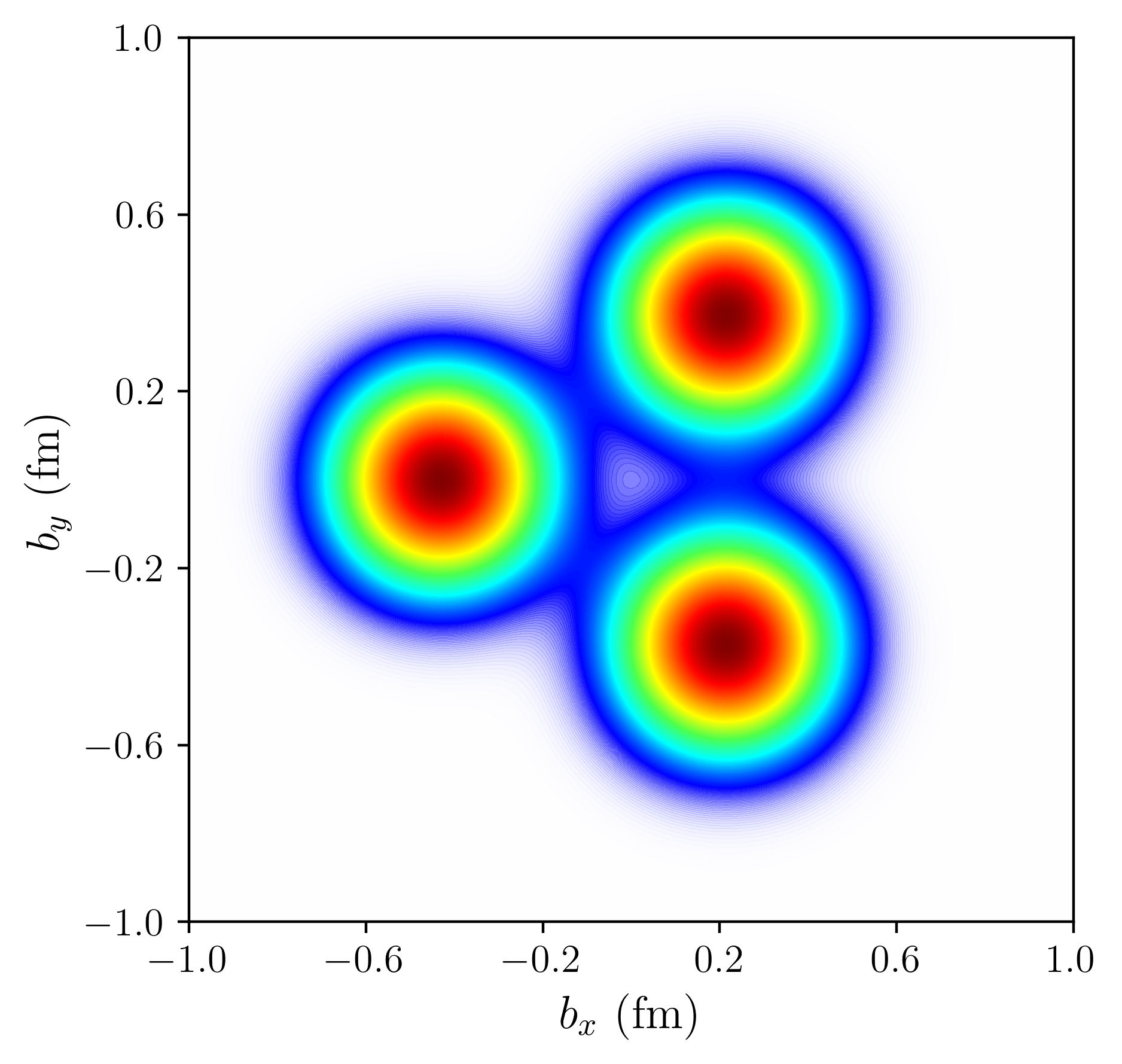}
    \caption{}
\end{subfigure}\hfill
\begin{subfigure}{0.355\textwidth}
    \centering
    \includegraphics[width=\textwidth]{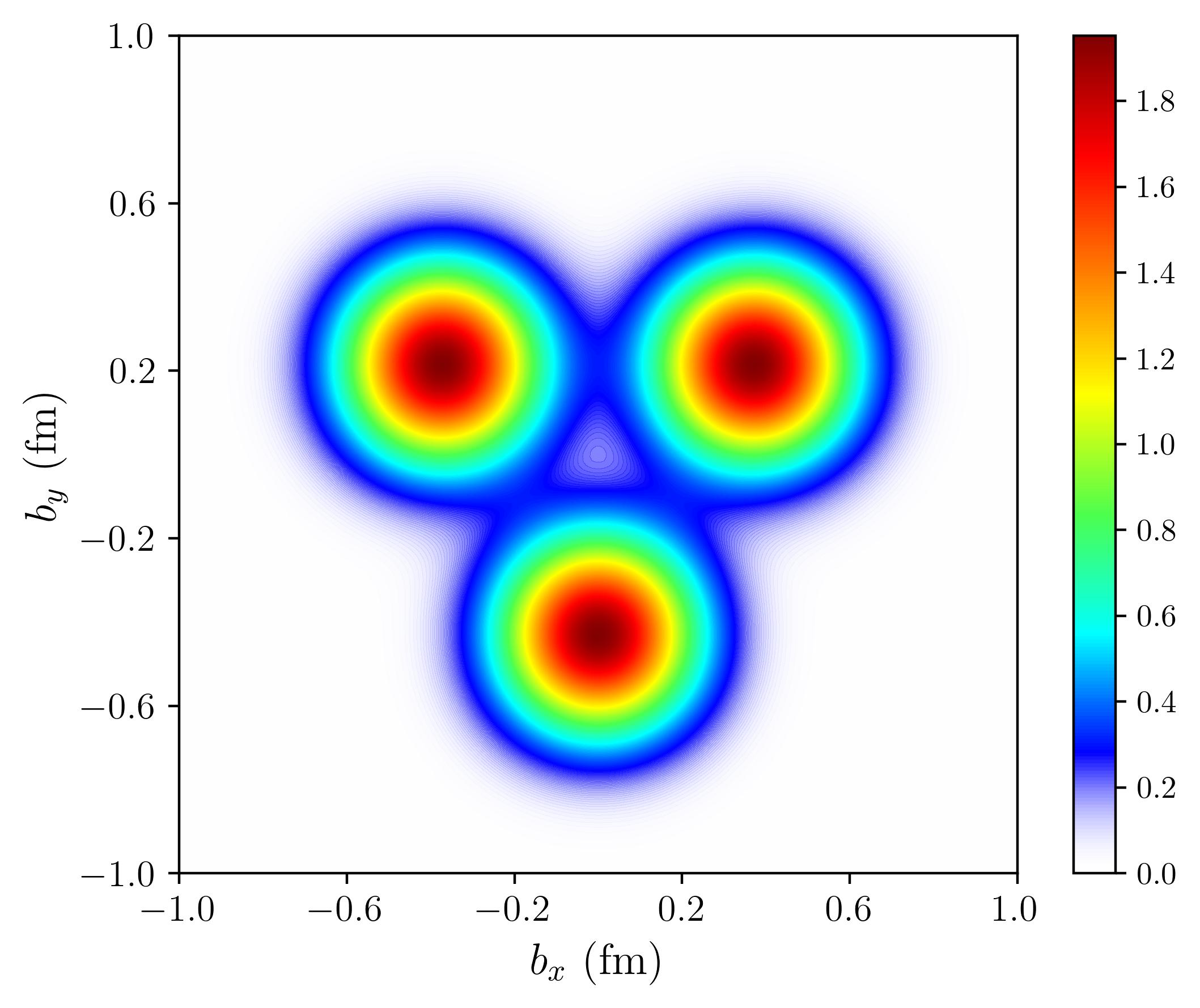}
    \caption{}
\end{subfigure}

\caption{Example contour plots of the transverse color-charge distribution at (a) $\theta=0$, (b) $\theta=\frac{\pi}{6}$, and (c) $\theta=\frac{\pi}{3}$ using $b_0=0.43$ fm and $R_p=0.165$ fm.}
\label{rotationcontourplot}
\end{figure}

\begin{figure}[h!]
\centering

\begin{subfigure}{0.48\textwidth}
    \centering
    \includegraphics[width=\textwidth]{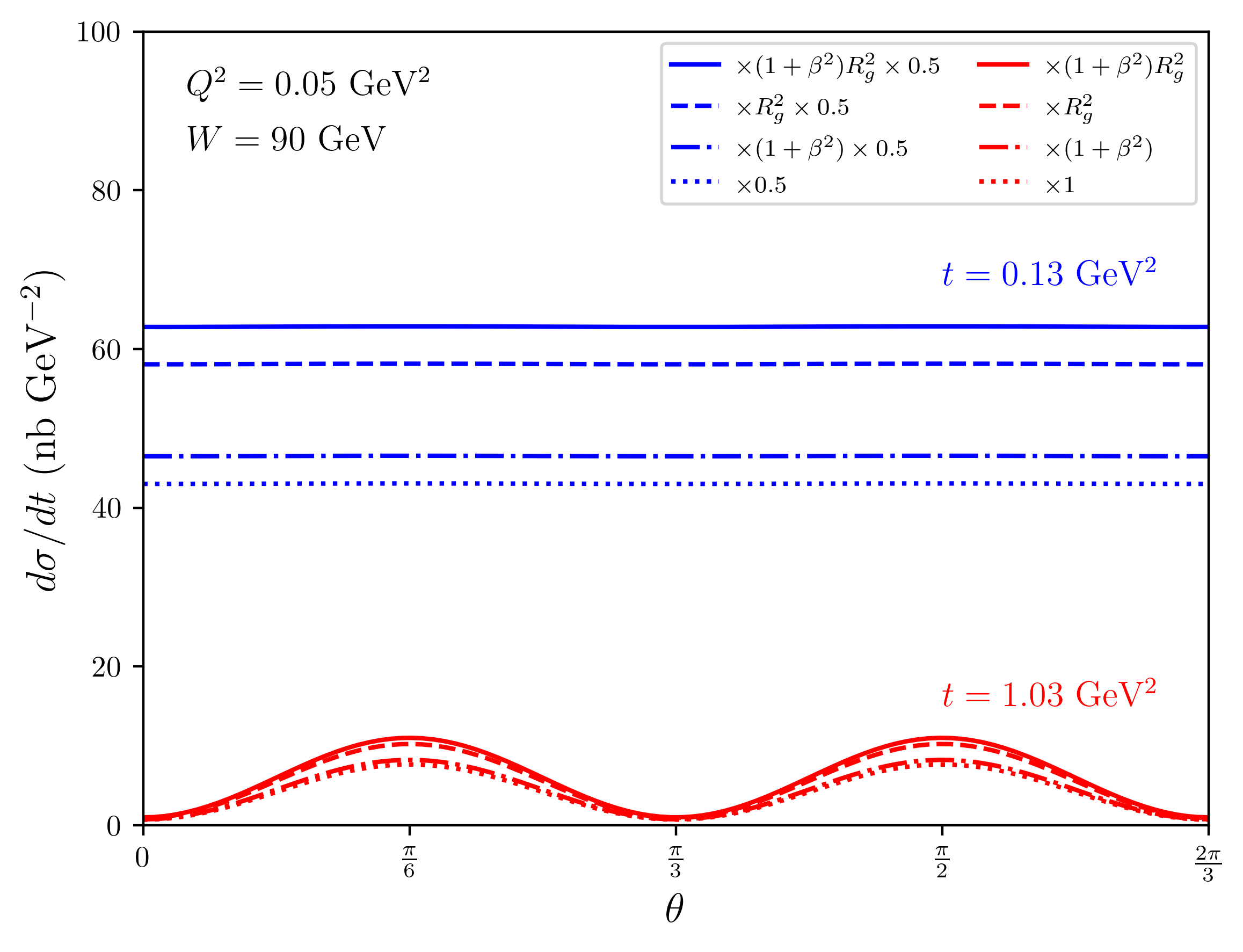} 
    \caption{}
\end{subfigure}
\hfill
\begin{subfigure}{0.48\textwidth}
    \centering
    \includegraphics[width=\textwidth]{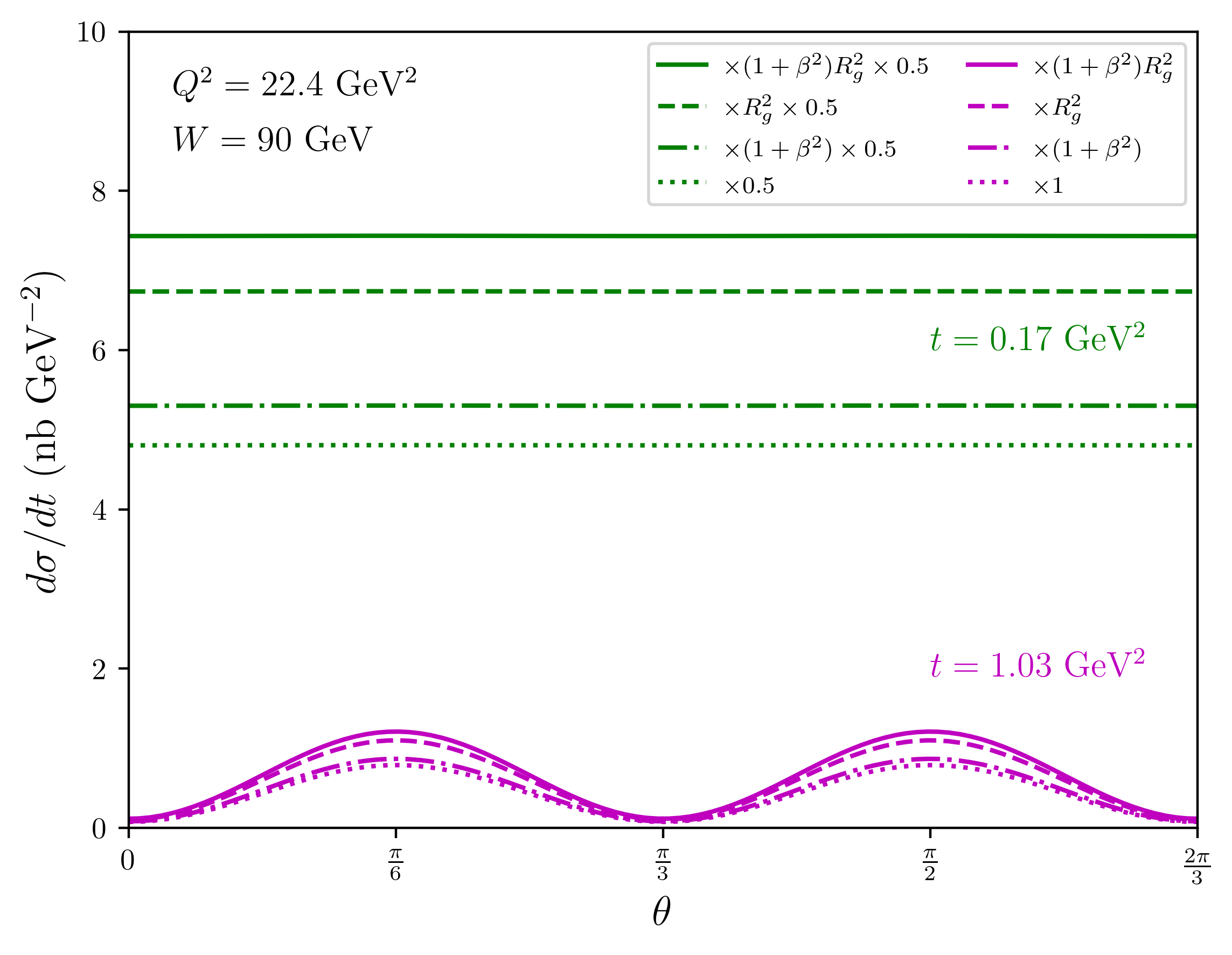}
    \caption{}
\end{subfigure}

\caption{The predicted value of differential cross section as a function of $\theta$, together with the comparison between each correction term applied in the small- and large-$t$, are shown for (a) $Q^2=0.05$ GeV$^2$ and (b) $Q^2=22.4$ GeV$^2$.}
\label{diffcrosssecvstheta}
\end{figure}

In Fig.~\ref{rotationcontourplot} we show contour plots of the proton profile in the transverse plane, along with the applied rotations of the valence quarks. The rotations are taken to be clockwise. Fig.~\ref{diffcrosssecvstheta} illustrates the effect of rotating the valence quarks on the differential cross section for the minimum and maximum values of both $t$ and $Q^2$ considered in this study, which lie within the HERA kinematic range. For both photoproduction (Fig.~\ref{diffcrosssecvstheta}a) and electroproduction (Fig.~\ref{diffcrosssecvstheta}b), the differential cross section exhibits sinusoidal fluctuations with a period of $\pi/3$. This reflects a symmetry of the rotation, such that the same pattern is obtained when the rotation direction is reversed (counterclockwise). In all cases, both $\left|\mathcal{A}^{\gamma^{(*)}p\rightarrow Vp}_{T,L}\right|^2$ and the correction terms vary with $\theta$; the effect of the rotation is enhanced with increasing $t$. 

To characterize the sensitivity of the cross section to rotational configuration, we show a band in the differential cross-section calculation representing the change in its magnitude due to rotation of the valence-quark configuration. The band is defined as the region bounded by the minimum and maximum predicted differential cross sections obtained when $\theta$ is varied over all possible angles. Fig.~\ref{diffcrosssecvstheta} further reveals a repeating rotational pattern: the angle configurations $\theta=(0.12+\frac{k\pi}{3})$, with $k=0,1,2,\dots,5$, yield identical differential cross sections, reproducing the same description as the initial fit.

\begin{figure}[h!]
\centering

\begin{subfigure}{0.48\textwidth}
    \centering
    \includegraphics[width=\textwidth]{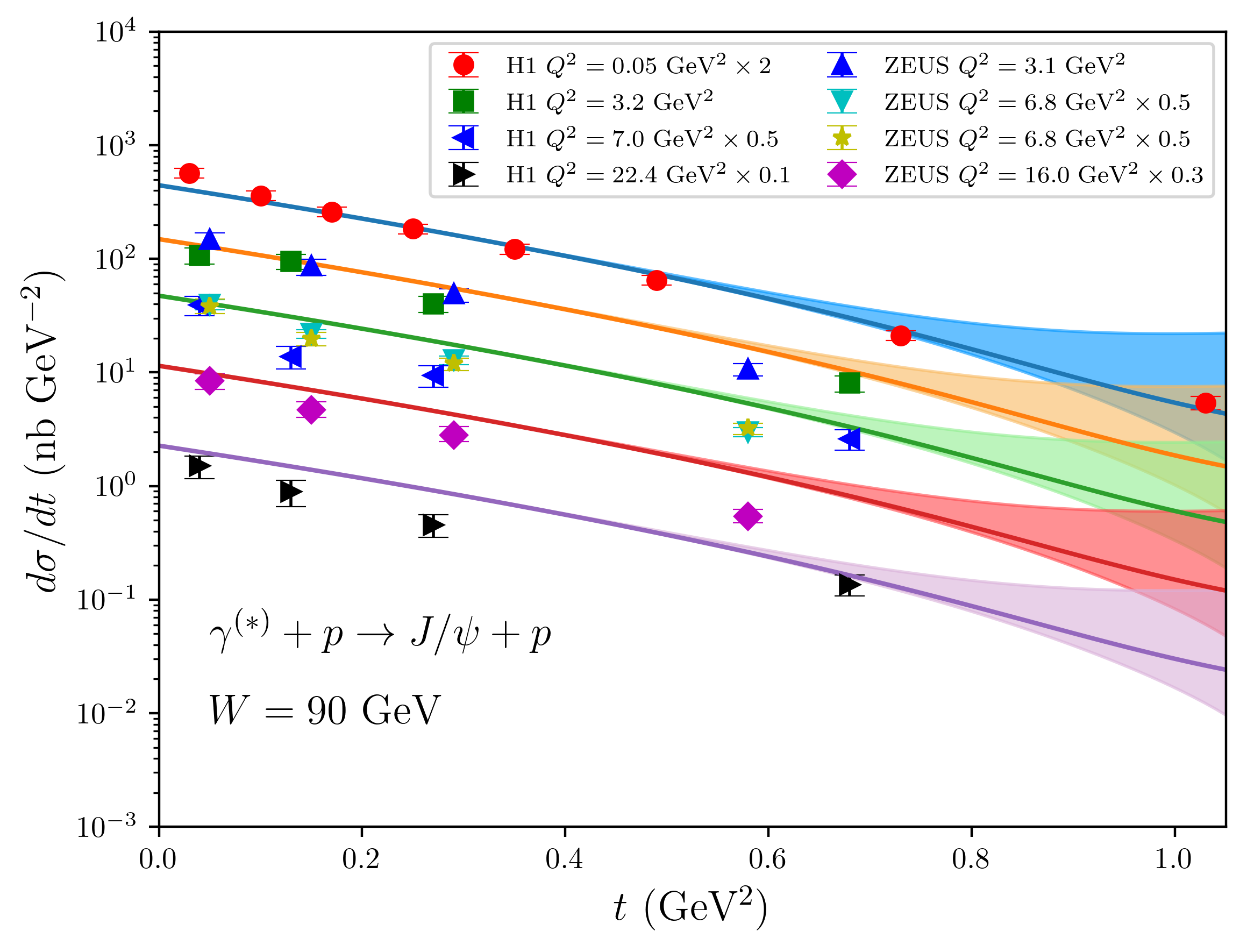}
    \caption{}
\end{subfigure}
\hfill
\begin{subfigure}{0.48\textwidth}
    \centering
    \includegraphics[width=\textwidth]{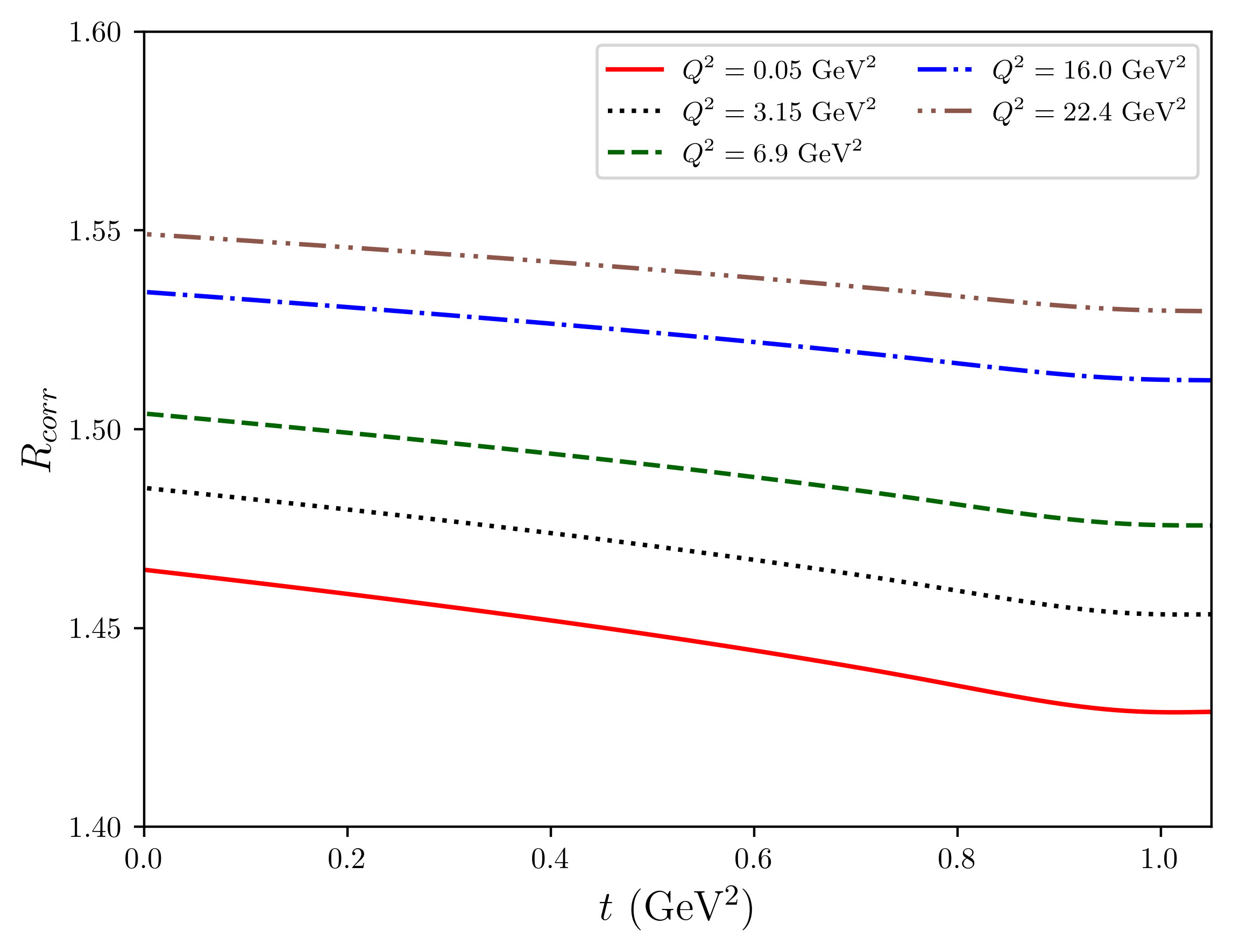}
    \caption{}
\end{subfigure}

\caption{(a) Model fit of the exclusive coherent $J/\psi$ productions as a function of $t$. 
The shaded bands represent the range of differential cross-section amplitudes obtained from all possible azimuthal rotation configurations of the valence quarks. 
The data are obtained from H1 \cite{H1:2005dtp} and ZEUS \cite{ZEUS:2004yeh} experiments. 
(b) The variable $R_{\text{corr}}$ represents the ratio between two schemes for calculating the differential cross section, as defined in Eq. \eqref{rcorr}.}
\label{modelfittdep}
\end{figure}

\begin{figure}[h]
\centering

\begin{subfigure}{0.48\textwidth}
    \centering
    \includegraphics[width=\textwidth]{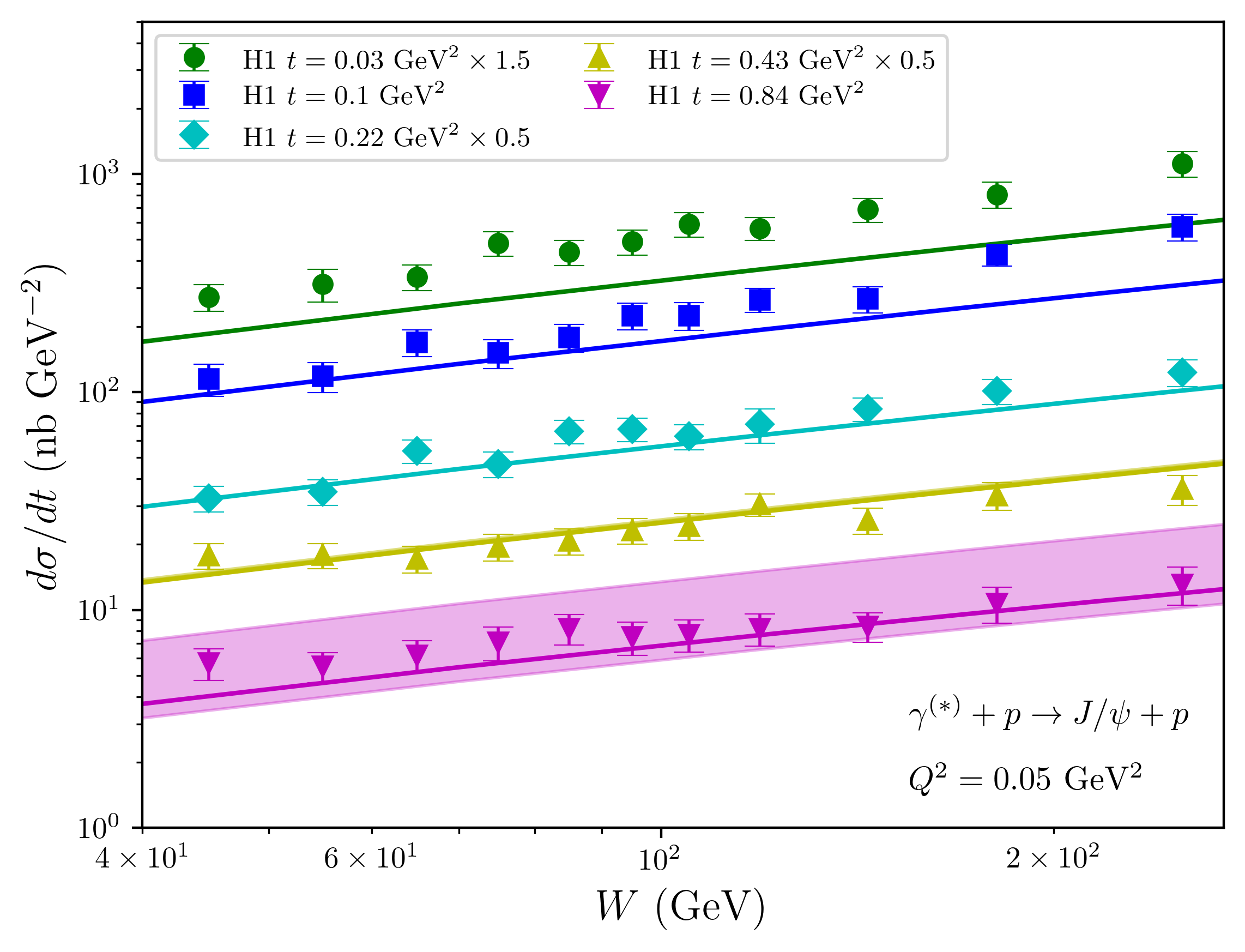} 
    \caption{}
\end{subfigure}
\hfill
\begin{subfigure}{0.48\textwidth}
    \centering
    \includegraphics[width=\textwidth]{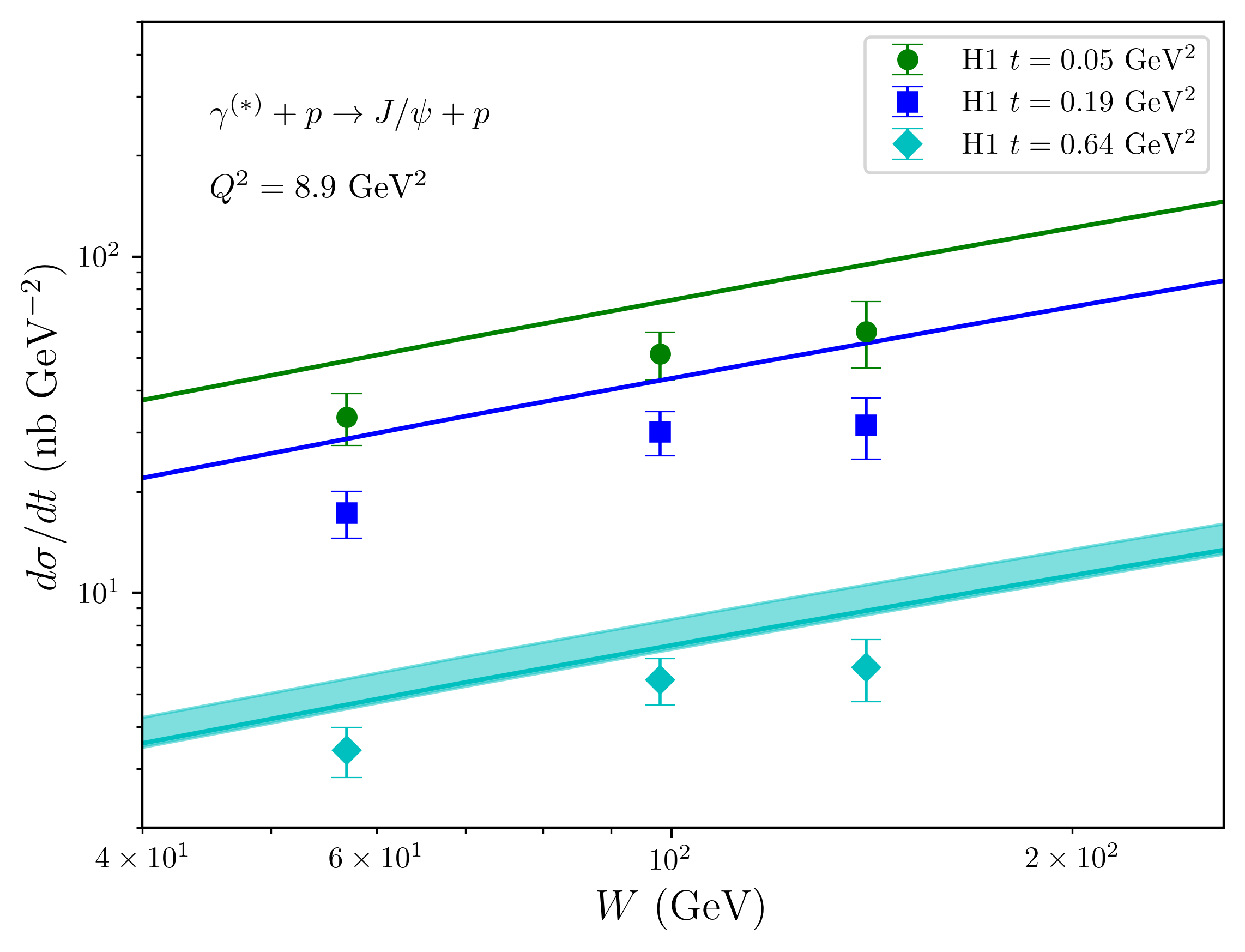}
    \caption{}
\end{subfigure}

\caption{The $W$ dependence of the model fit is compared with the H1 coherent $J/\psi$ productions data for (a) $Q^2=0.05$ GeV$^2$ and (b) $Q^2=8.9$ GeV$^2$ with the same parameters as in Fig.~\ref{modelfittdep}. Bands corresponding to all rotational configurations are also included.}
\label{modelfitwdep}
\end{figure}

In Fig.~\ref{modelfittdep} we present the description of photoproduction and electroproduction of the $J/\psi$, including the corresponding bands.
In the photoproduction case, the model tends to underestimate the differential cross section at very low $t$. Moreover, a mild shoulder is visible in the region $0.2\le t \le 0.7$ GeV$^2$, followed by a steeper fall-off at large $t$. The considerable change in slope at large $t$ is controlled by the rotation angle, with $\theta=\frac{k\pi}{3}$ give the steepest slope, while those with $\theta=(\frac{\pi}{6}+\frac{k\pi}{3})$ produce the flattest behavior. The differential cross section bands become considerably broader in the region $t \ge 0.5$ GeV$^2$. The rotational effect shows only a weak dependence on $Q^2$. In particular, the relative difference of the differential cross section with respect to the best-fit configuration is only on the order of $8\%$ between photo- and electroproduction at $Q^2=22.4$ GeV$^2$.

Across the considered $Q^2$ range, the combined phenomenological correction increases gradually with $Q^2$. In this kinematical region, the skewedness correction reaches up to about $40\%$, while the real-part correction is at most $11\%$. The combined correction shows only a weak dependence on $t$. Based on this observation, the fit was performed without including the phenomenological corrections. Comparison with the $t$-slope obtained from the fit including corrections shows almost no visible difference. The relative change in the cross section is $0\%$ at $Q^2=0.05$ GeV$^2$ and reaches at most $1\%$ at $Q^2=22.4$ GeV$^2$. The difference in slope is approximately $0.1\%$, with the fit without corrections being slightly steeper. Thus, the correction factors effectively act as a free parameter, analogous to $\bar{\chi}$, which controls the magnitude of the differential cross section. Their net effect can therefore be absorbed by an increase in $\bar{\chi}$.

In Fig.~\ref{modelfitwdep}, we also show the $W$ dependence obtained using the parameters from the $t$-dependent fit. The results indicate that the proton profile does not affect the $W$-slope of the differential cross section. Instead, it purely reflects how we matched the expected magnitude in the initial fit. The large deviation at $t\le0.1$ GeV$^2$ in photoproduction (Fig~\ref{modelfitwdep}a) originates from the underestimation of the data in our initial $t$-dependent fit. A similar deviation at $Q^2=8.9$ GeV$^2$ (Fig~\ref{modelfitwdep}b) is caused by an overestimation of the electroproduction data. Based on $W$ dependence fit, we can further verify that the bands affect only the $t$ dependence and show no dependence on $W$. This is indicated by Fig.~\ref{modelfitwdep}, where the separation between the upper and lower edges of the bands remains constant for any value of $W$ at a given $t$.

\begin{figure}[h]
\centering

\begin{subfigure}{0.48\textwidth}
    \centering
    \includegraphics[width=\textwidth]{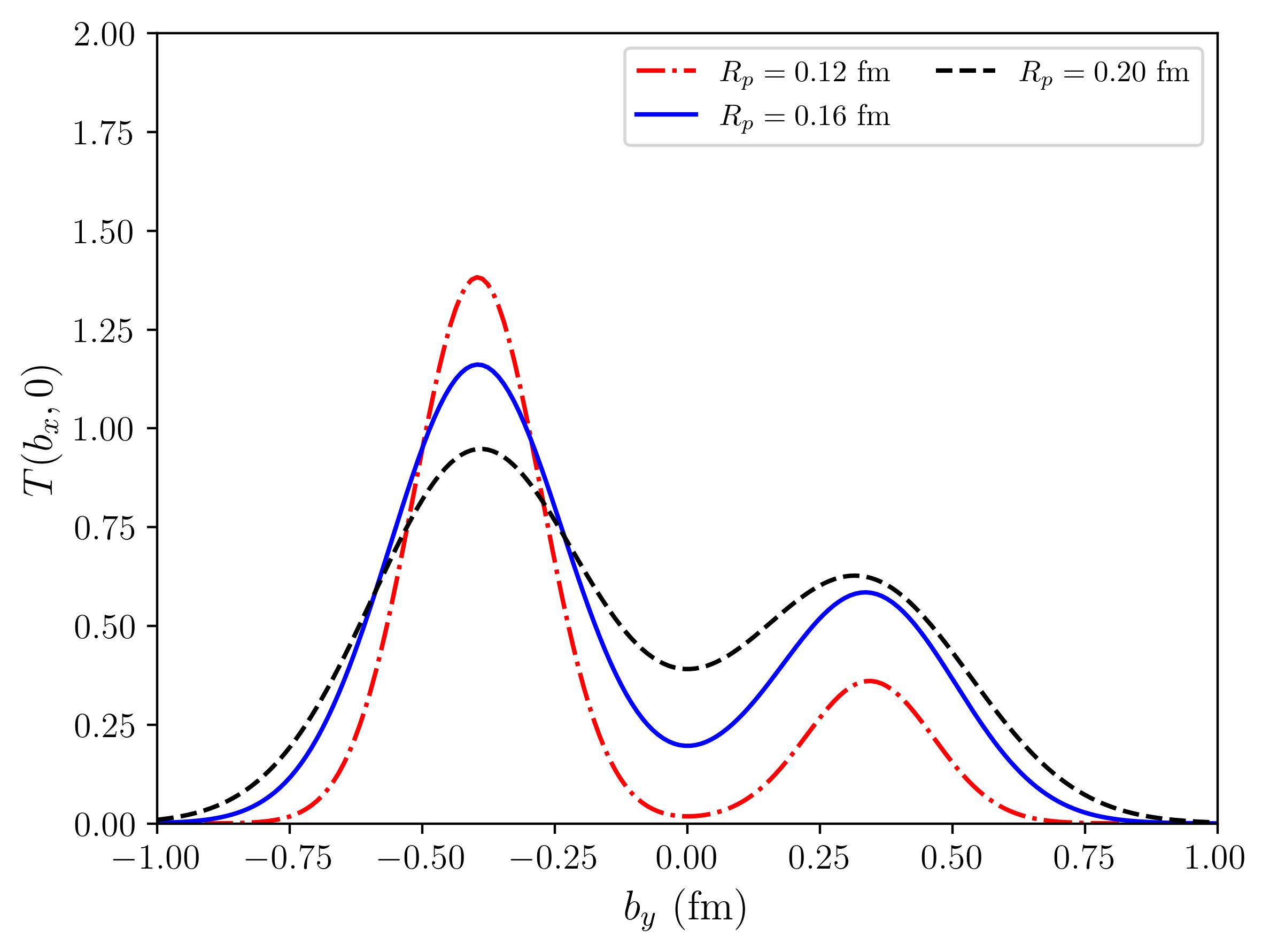} 
    \caption{}
\end{subfigure}
\hfill
\begin{subfigure}{0.48\textwidth}
    \centering
    \includegraphics[width=\textwidth]{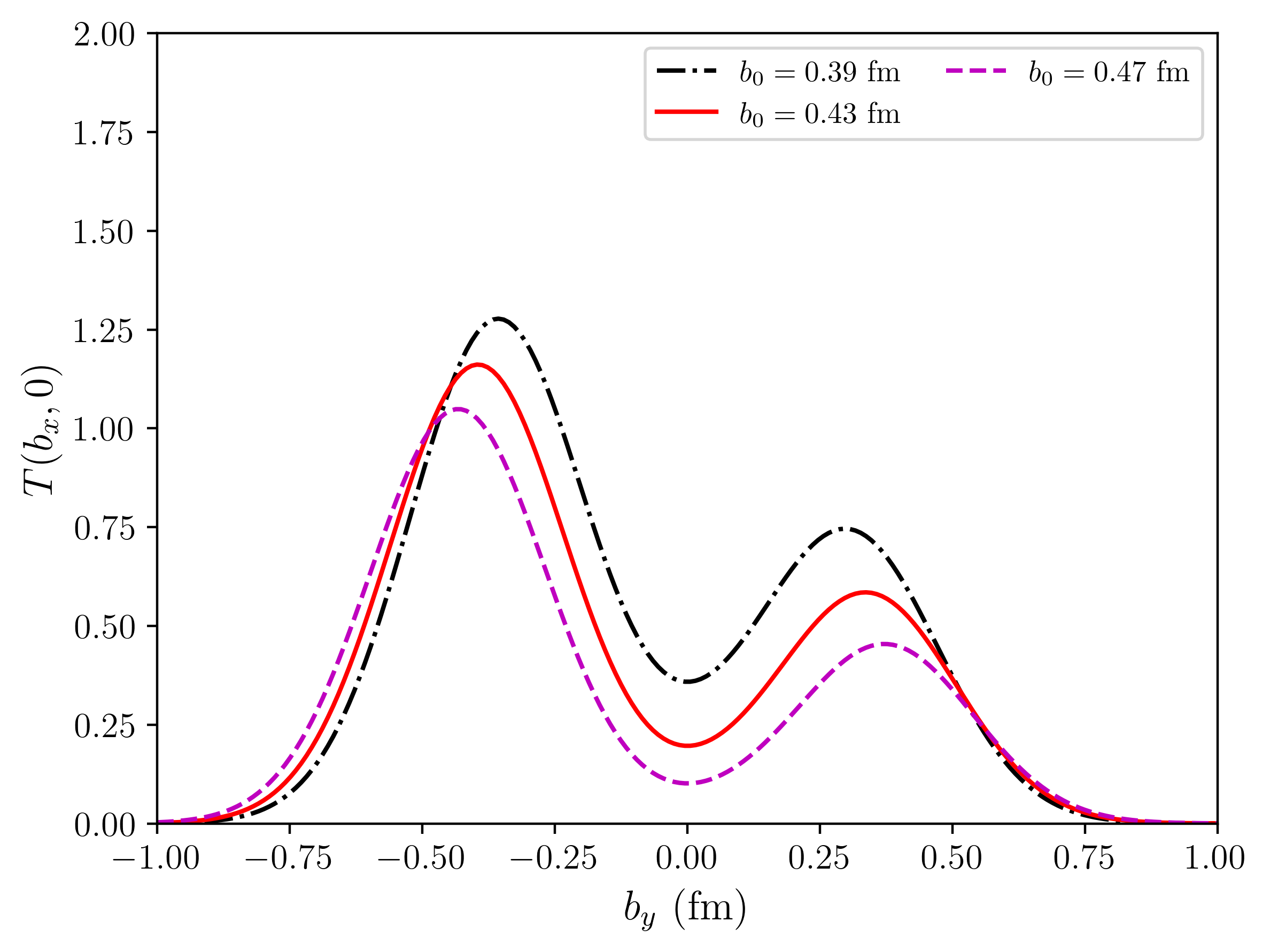}
    \caption{}
\end{subfigure}

\vspace{0.5em} 

\begin{subfigure}{0.48\textwidth}
    \centering
    \includegraphics[width=\textwidth]{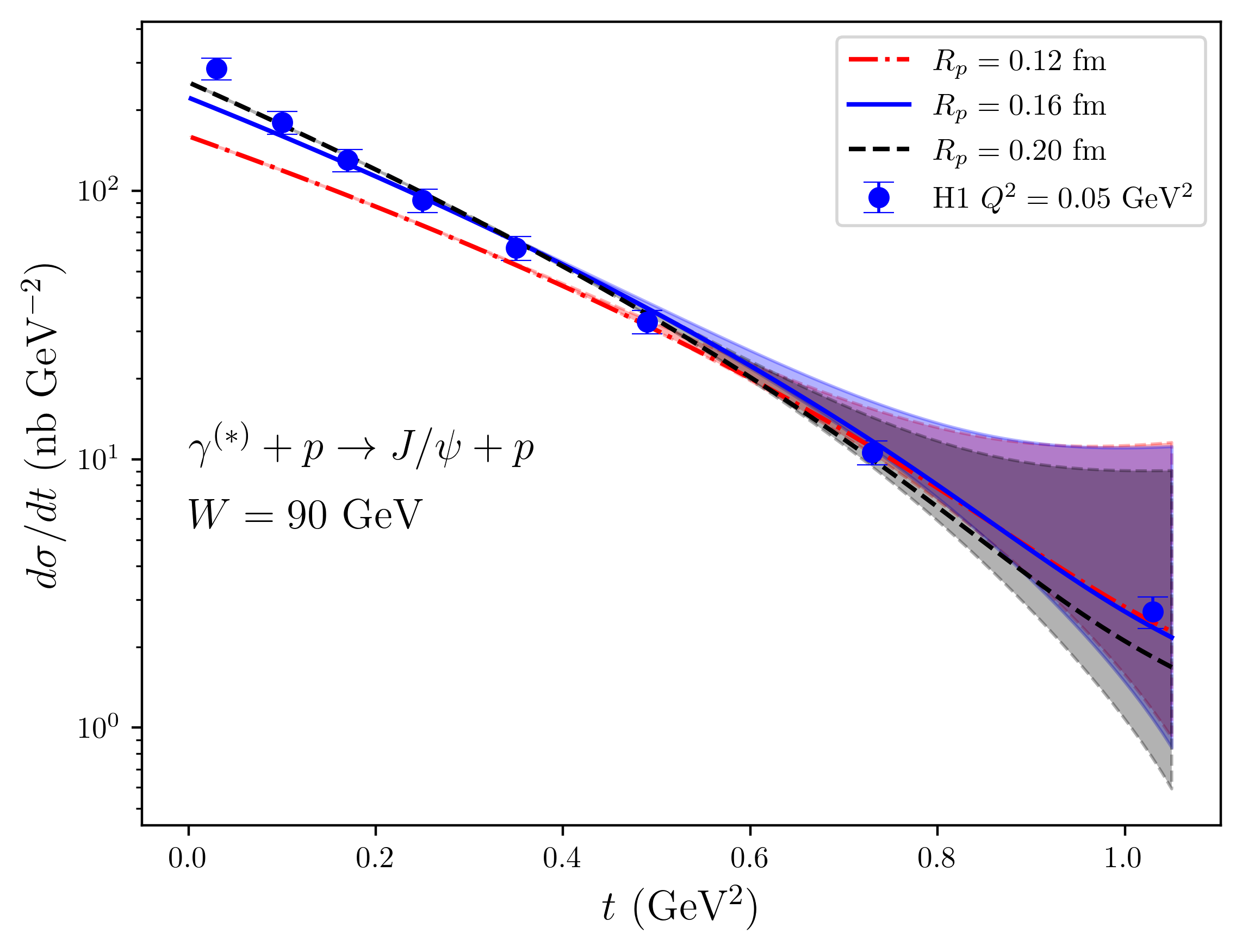} 
    \caption{}
\end{subfigure}
\hfill
\begin{subfigure}{0.48\textwidth}
    \centering
    \includegraphics[width=\textwidth]{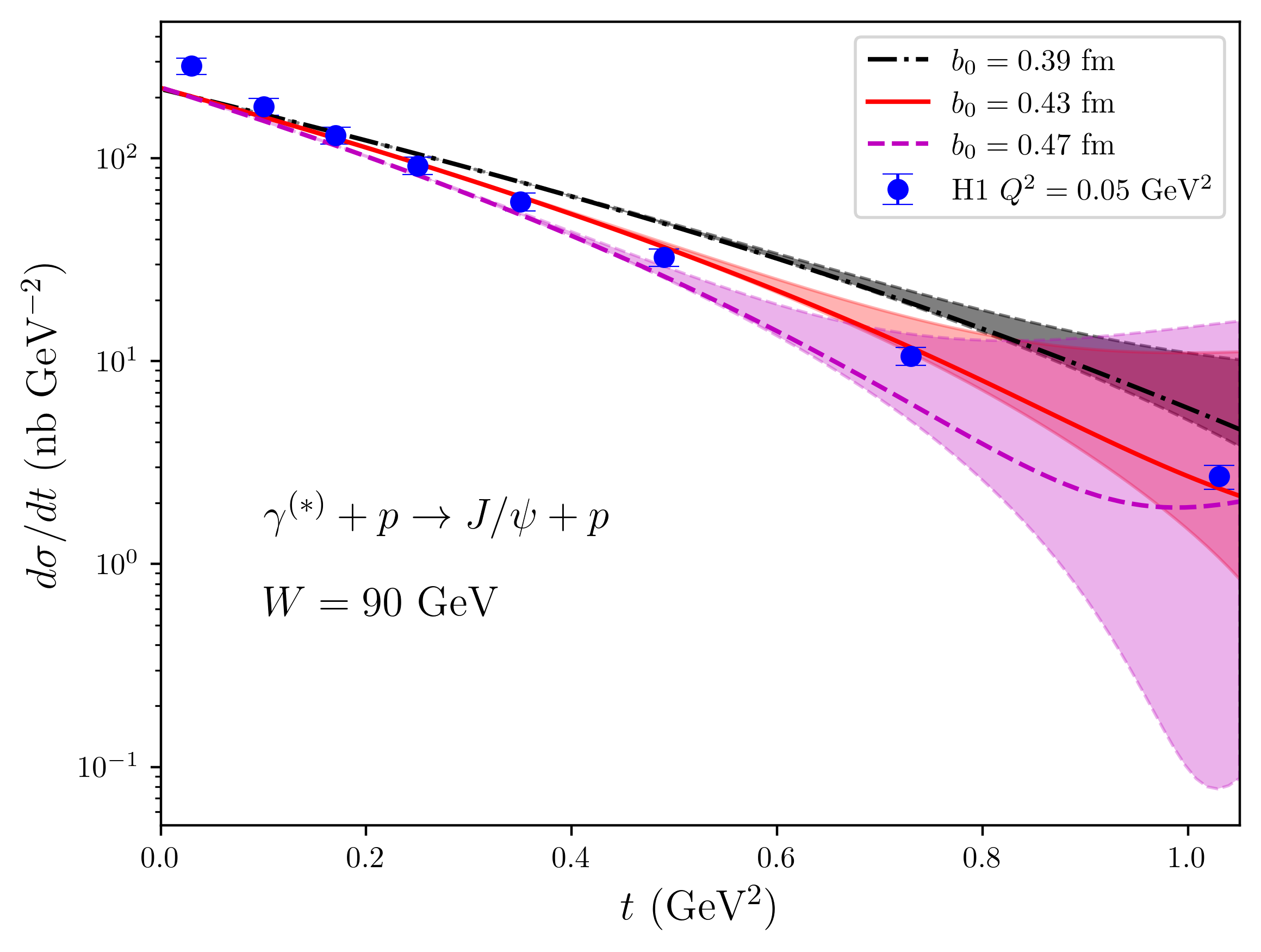}
    \caption{}
\end{subfigure}

\caption{Top: Comparison of the spatial color-charge distributions of the proton for variations in (a) $R_p$ and (b) $b_0$. Bottom: Description of the coherent $J/\psi$ photoproduction data for different values of (c) $R_p$ and (d) $b_0$ corresponding to each parameter variation, together with bands representing all possible rotational angle configurations.}
\label{rpb0var}
\end{figure}

We now investigate the effect of varying $b_0$ and $R_p$ on the description of the photoproduction data. From the proton profiles in Figs.~\ref{rpb0var}a and \ref{rpb0var}b, we see that both increasing $b_0$ and decreasing $R_p$ reduce the overlap of the color-charge distributions near $(b_x,b_y)=(0,0)$. In other words, the Gaussian profiles of the three valence quarks become more separated. This feature leads to a steeper $t$-slope in the predicted cross section. The slope is much more sensitive to changes in $b_0$ than to changes in $R_p$. Although several parameter choices appear to provide a better slope pattern compared to the data, we do not combine these variations in the fit because they would violate the condition $b_0 + R_p \leq 0.6$ fm, thereby implying an unrealistically large proton size.

The variations of $b_0$ and $R_p$ also affect the bands associated with the valence-quark rotational configurations. A larger overlap region among the Gaussian profiles of the hot spots, achieved for smaller $b_0$ or larger $R_p$, leads to narrower bands.  For changes in $R_p$, the difference in the amplitude is relatively small, at the level of about $2\%$–$8\%$. In contrast, the variation of $b_0$ has a much stronger effect. The largest band is obtained at $b_0 = 0.47$ fm, where the model even predicts the emergence of a dip at $t = 1.02$ GeV$^2$ for configurations with $\theta = \frac{k\pi}{3}$.

\section{Summary and Outlook} \label{conclusions}

In this work, we have applied a simple static Gaussian hot spot model to describe the proton’s transverse color-charge distribution. In this picture, the three valence quarks at large~$x$ act as localized sources of the small-$x$ parton sea, and thus each valence quark serves as the center of a hot spot. We tested this assumption by calculating the differential cross section for exclusive coherent $J/\psi$ production in the region $t<1.2\,\mathrm{GeV}^{2}$ and by comparing our results with HERA data. By introducing three geometric degrees of freedom in the hot spot configuration: the Gaussian width parameter of each hot spot, the radial distance of the valence quarks from the proton center in the transverse plane, and their azimuthal orientation, we examined how each parameter influences the behavior of the cross section. In particular, we quantified their respective impacts on both the exponential slope and the amplitude of the differential cross section.

We found that each degree of freedom affects the exponential slope with varying sensitivities. As expected, over the full range of~$t$ studied, the cross section is highly sensitive to the Gaussian width of the hot spots and to their radial location. The cross section at small~$t$, particularly for $t<0.5\,\mathrm{GeV}^{2}$, shows only weak sensitivity to the rotational degree of freedom. At larger~$t$, however, the orientation of the valence-quark configuration becomes increasingly important, leading to variations in the cross-section amplitude of up to two orders of magnitude at $t=1.2\,\mathrm{GeV}^{2}$ and thereby influencing the effective slope of the predicted distribution. This rotational effect is essentially independent of the photon virtuality. Although we included the phenomenological real-part and skewedness corrections in the fitting procedure, we demonstrated that these corrections primarily modify the overall amplitude of the differential cross section, with a combined effect of up to about $55\%$, decreasing slightly toward larger~$t$, while having only a minor influence on the exponential slope.

This study provides guidance for the modeling of Gaussian-like proton shapes, particularly regarding the geometric orientation of the valence-quark configuration, which may be connected to the dipole orientation induced by the probing photon. In the future, it will be important to test the universality of the geometric sensitivity extracted here by examining the exclusive production of other vector mesons, as well as incoherent processes in which the hot spot geometry and its fluctuations are expected to play an even more significant role. One may also investigate how different choices of the vector-meson wave function influence this proton geometrical configuration sensitivity. Finally, since the angle-dependent position of the hot spot will affect the calculated cross section, it would be interesting to investigate whether incorporating intrinsic orbital motion of the valence quarks can also affect the cross-section calculation.

We also note that in this work we adopted a simple equilateral-triangle arrangement for the hot spot positions, thereby neglecting possible effects associated with the different charges of the valence quarks, which could in principle shift their relative positions. Investigating such charge-dependent geometric variations would be one of our most promising directions for future studies.

\end{document}